\documentclass[a4paper,twocolumn,twoside,10pt]{article}
\usepackage{graphicx,amsmath,amssymb,amsthm}
\usepackage{siunitx}
\usepackage[numbers]{natbib}

\makeatletter \oddsidemargin-.25in \evensidemargin-.25in
\makeatother \topmargin-0.75in \textwidth7in \textheight9.5in

\graphicspath{ {figures/} }
\usepackage[caption=false]{subfig}
\usepackage{multirow}

\theoremstyle{definition}

\newcommand{\AIC}{\text{AIC}}

\begin{document}
\def \thepage {}
\date{}

\title{\begin{flushleft}
\noindent {\small {\it Proceedings of ICCSA 2014
\\Normandie University, Le Havre, France - June 23-26, 2014\\[5mm]
 }}
\end{flushleft}\Large\bf \uppercase{Dynamics of Media Attention} }

\author{V.A. Traag\thanks{Corresponding author: 
  \texttt{vincent@traag.net}}, 
  R. Reinanda, J. Hicks, G. van Klinken
  \thanks{All authors are with the KITLV, Leiden, the Netherlands}
  \thanks{R. Reinanda is associated to Faculty of Sciences, University of
  Amsterdam, the Netherlands}}
 \maketitle


{\footnotesize \noindent {\bf Abstract.}  
Studies of human attention dynamics analyses how attention is focused on
specific topics, issues or people. In online social media, there are clear signs
of exogenous shocks, bursty dynamics, and an exponential or powerlaw lifetime
distribution.  We here analyse the attention dynamics of traditional media,
focussing on co-occurrence of people in newspaper articles. The results are
quite different from online social networks and attention. Different regimes
seem to be operating at two different time scales. At short time scales we see
evidence of bursty dynamics and fast decaying edge lifetimes and attention. This
behaviour disappears for longer time scales, and in that regime we find
Poissonian dynamics and slower decaying lifetimes. We propose that a cascading
Poisson process may take place, with issues arising at a constant rate over a
long time scale, and faster dynamics at a shorter time scale.

}

\vskip.2in


\section{Introduction}

With the arrival of large scale data sets, interest in quantifying human
attention rose. It became possible to measure precisely how attention grew
and decayed~\cite{crane_robust_2008}.  Moreover, it appeared that many human
dynamics showed signs of bursty behaviour: short windows of intense activity
with long intermittent time spans of inactivity~\cite{malmgren_poissonian_2008}.
The duration a person is active---the time between its first and last
occurrence, i.e. its lifetime---seems to decay as an exponential, while the edge
lifetime seems to follow a powerlaw
distribution~\cite{hidalgo_dynamics_2008,leskovec_microscopic_2008}.

We analyse a large dataset of newspaper articles from traditional printed media.
We show that the dynamics of this dataset are different from social media.  Our
data consist of $140\,263$ newspaper articles from Indonesia from roughly 2004
to 2012, gathered by a news service called \emph{Joyo}, mainly focussing on
political news. We automatically identify entities by using a technique known as
named entity recognition, and only retain person names that occur in more
documents than on average (we discard organisations and
locations)~\cite{finkel_incorporating_2005}. We then construct a network by
creating a node for every person and an edge for each co-occurrence between two
persons, and we record the date of the co-occurrence.  We only take into account
co-occurrences of people in the same sentence, and only about $3.2\%$ of the
sentences contain more than one person, so this is relatively restrictive. All
time is measured in days, denoted by the symbol $\si{\day}$, and we report
standard errors for estimations.

\section{Results}

\begin{figure}[t]
  \begin{center}
    \includegraphics[width=\columnwidth]{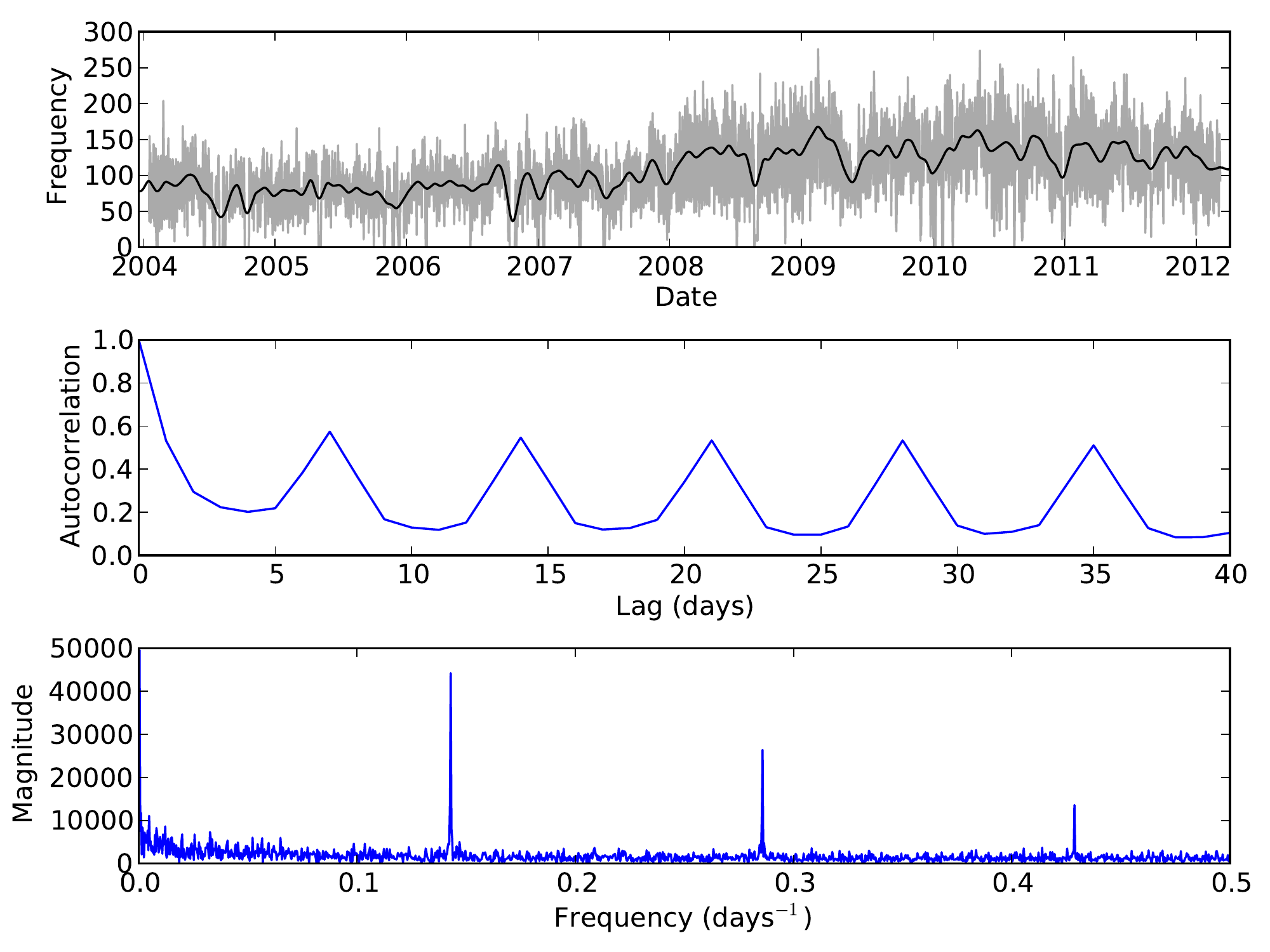}
  \end{center}
  \caption{Daily node frequency. The first panel contains the number of nodes in
    the newspaper (that have at least one co-occurrence) per day (in gray), with
    the solid black line representing a smoothing (using convolution with a Hann
    window of 8 weeks) of this daily frequency. The second panel displays the
    autocorrelation function, which shows a clear peak at a lag of $7$ days,
    indicating there is a weekly pattern in the data. This is also confirmed by
    Fourier analysis in the third panel, with a clear peak at a frequency
    of $\SI{1/7}{\day^{-1}} \approx \SI{0.14}{\day^{-1}}$.}
  \label{fig:daily_node_freq}
\end{figure}

\begin{figure}[t]
  \begin{center}
    \includegraphics[width=\columnwidth]{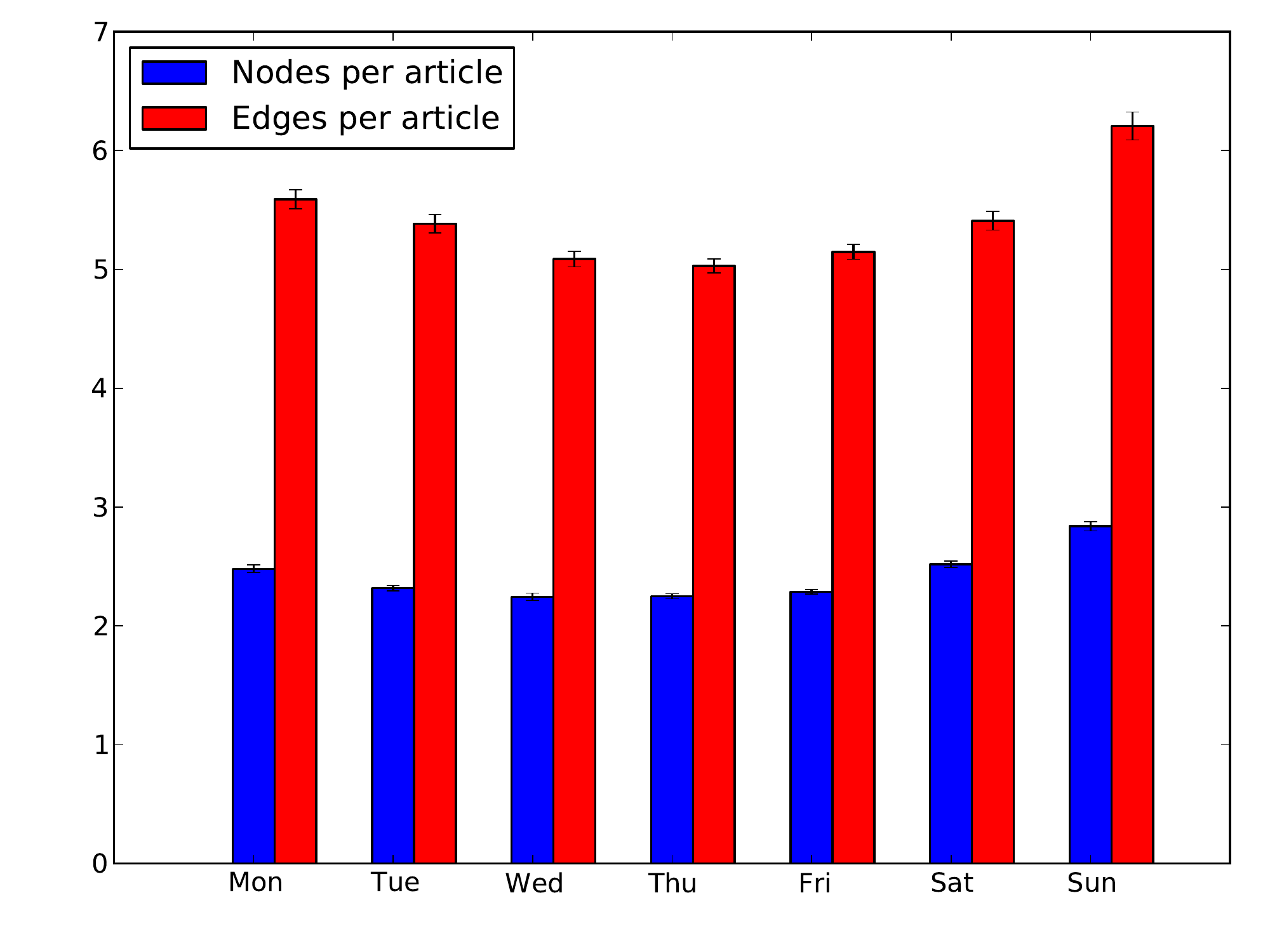}
  \end{center}
  \caption{Number of nodes and edges per article per day of the week. The number
    of nodes count how many nodes appear on average in an article. The number of
    edges count all the co-occurrences per article. Although the largest part
    of the cyclical behaviour is due to the weekly newscycle (weekend vs.
    weekday), there still remains a cyclical pattern after normalisation. 
}
  \label{fig:weekday_distr}
\end{figure}

In total, there are $n=9\,467$ nodes and they have about $\langle k_i \rangle
\approx 12$ neighbours on average. Two people co-occur on average about $3$
times. Let us first look at how these quantities vary over time. Let $E_t$ be
the number of co-occurrences at time $t$, and $N_t$ the number of nodes that
have at least one co-occurrence at time $t$. The dynamics of $N_t$ follow a
distinctive weekly pattern (Fig.~\ref{fig:daily_node_freq}). This is confirmed
by the autocorrelation function, which shows a clear peak at a lag of $7$ days
with a correlation of about $0.57$, while the Fourier transform has clear peaks
at a frequency of about $\SI{1/7}{\day^{-1}} \approx \SI{0.14}{\day^{-1}}$.
Results for $E_t$ follow a similar pattern. Although this largely coincides with
the weekly cycle of the number of articles, a cyclic pattern remains if the
frequencies are normalised by the article
frequencies~(Fig.~\ref{fig:weekday_distr}). For the nodes this pattern is weak,
but for the edges more noticeable. Surprisingly, the cycle attains its high
point at the end/beginning of the week, while its low point occurs in the middle
of the week.

\begin{figure*}[!tbh]
  \begin{center}
    \subfloat[Peak attention]{
      \includegraphics[width=0.48\textwidth]{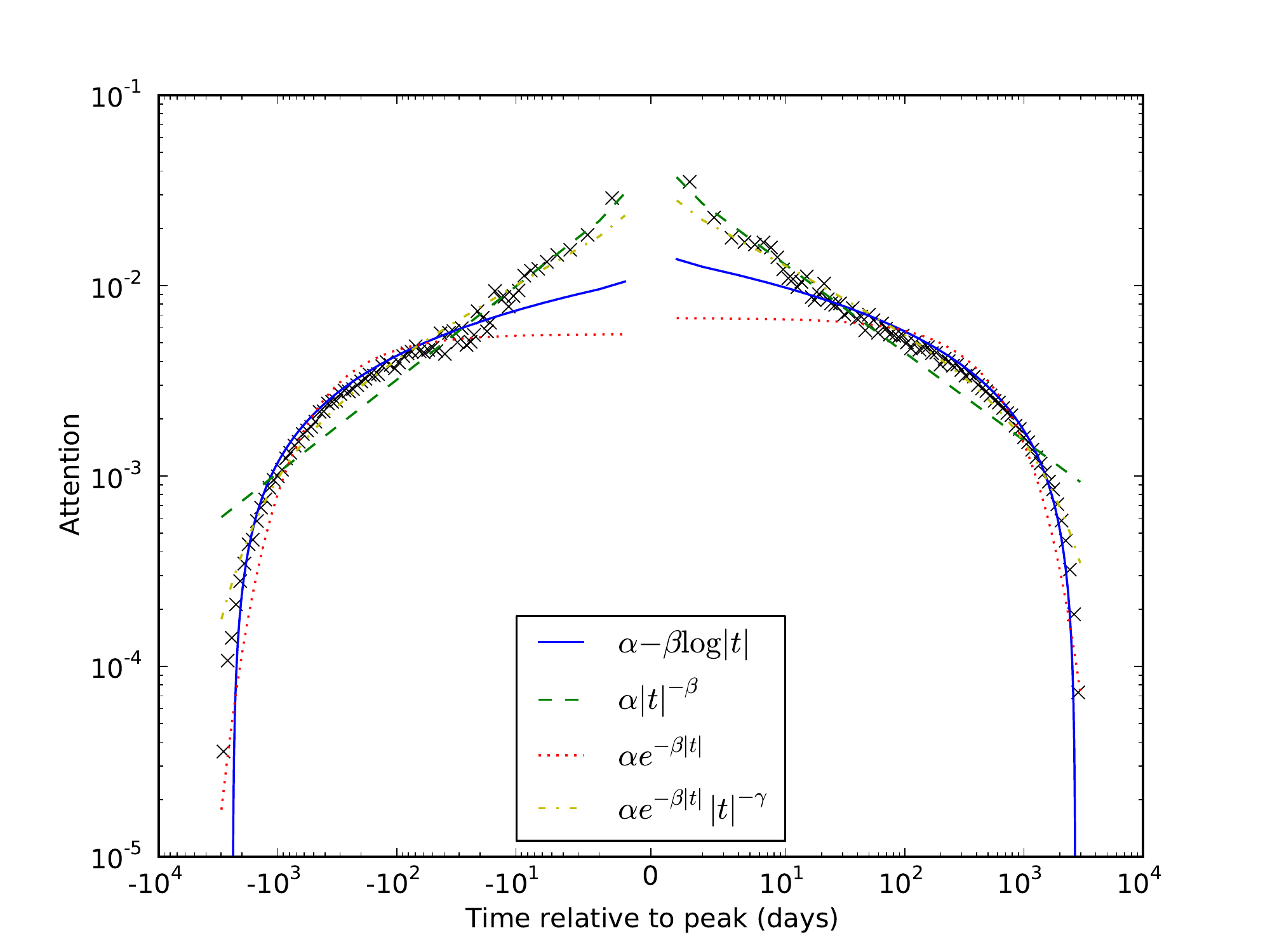}
      \label{fig:growth_decay_attention}
    }
    \subfloat[Inter event time]{
      \includegraphics[width=0.48\textwidth]{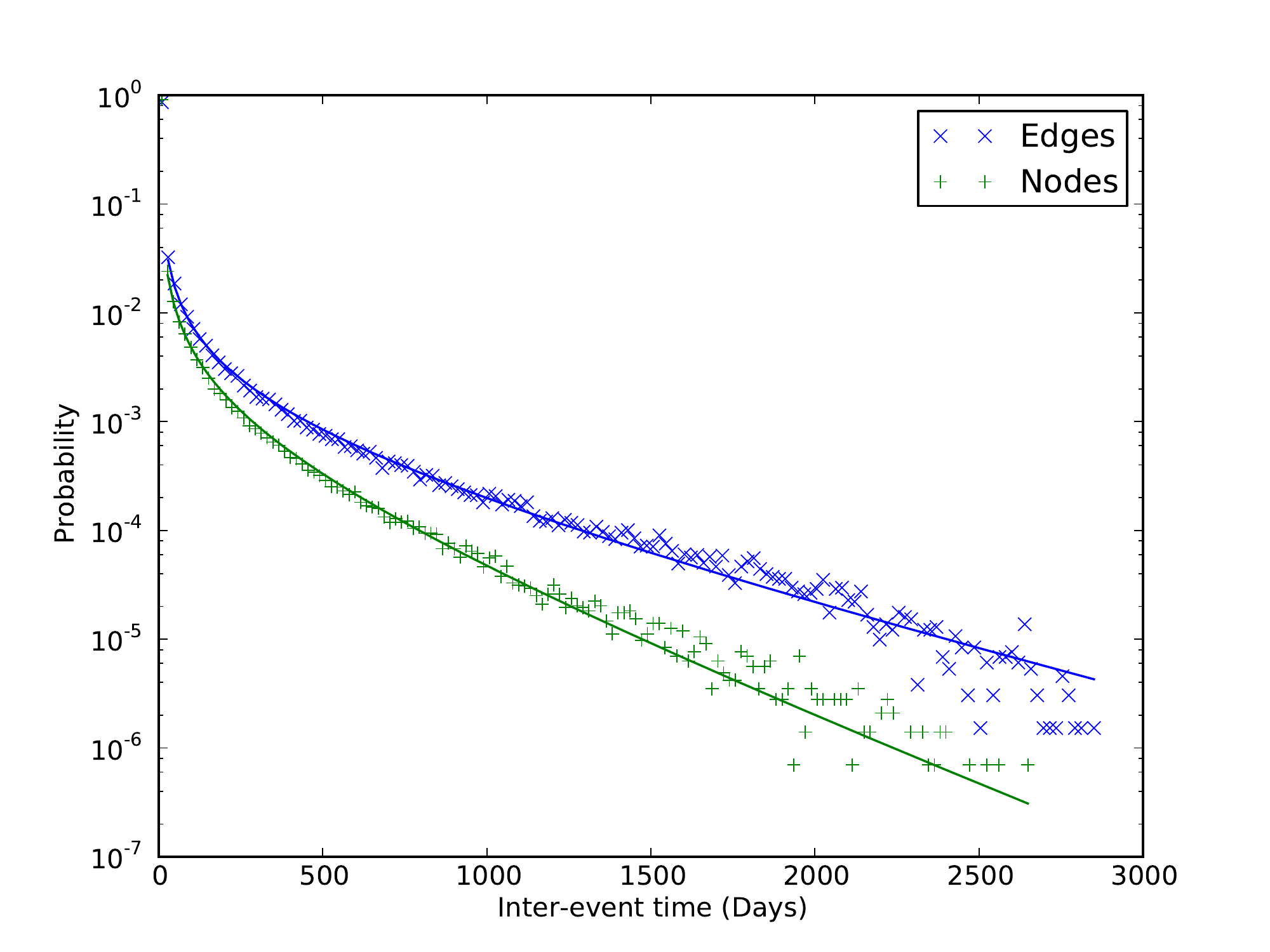}
      \label{fig:iets_distr}
    } \\
    \subfloat[Life time]{
      \includegraphics[width=0.48\textwidth]{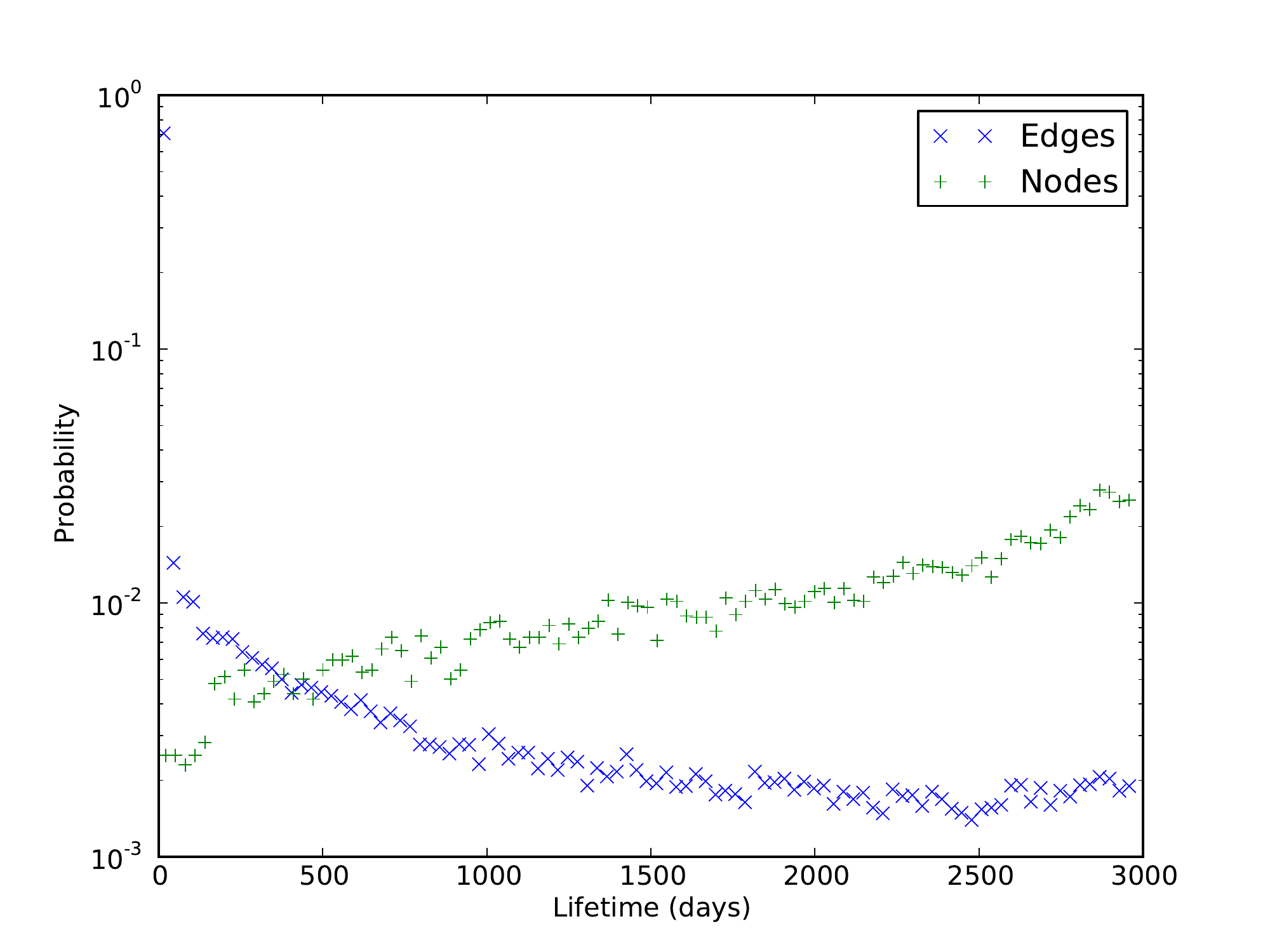}
      \label{fig:lifetime_distr}
    } 
    \subfloat[First time delay]{
      \includegraphics[width=0.48\textwidth]{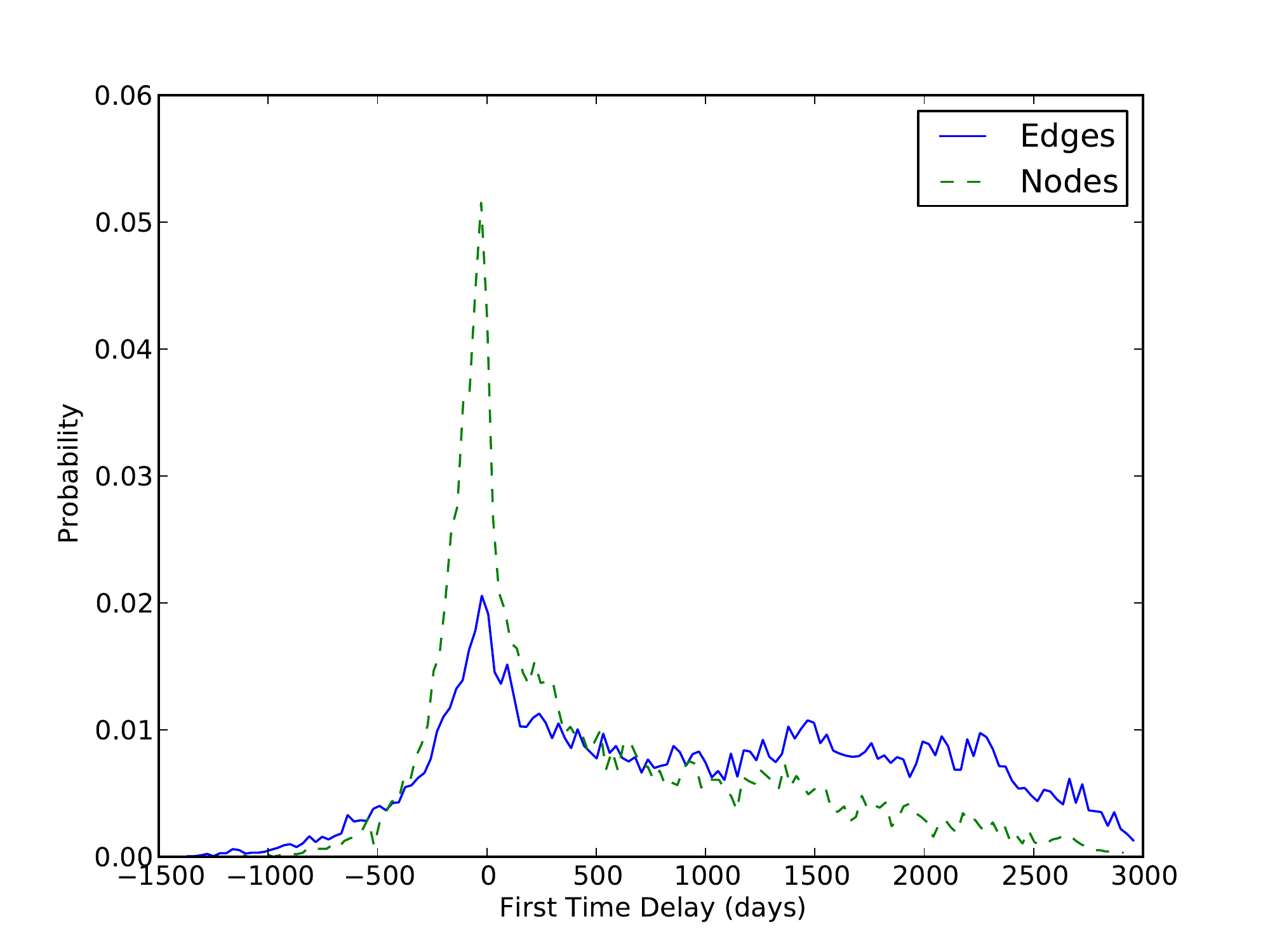}
      \label{fig:first_time_delay_distr}
    }
  \end{center}
  \caption{Attention statistics. Panel (a) displays how attention---as in the
    frequency of occurring---grows and decays with the peak attention centred at
    zero. Panel (b) shows the inter-event time distribution, which has a clear
    exponential tail, suggesting a Poisson process. Panel (c) features the
    distribution of the lifetime---the amount of time between the first and last
    time of observing that node or edge. The edge lifetime decays very fast for
    the first few days, but then decays much slower. The node lifetime follows
    an increasing distribution, indicating they can have a very long lifetime.
    Finally, panel (d) shows the distribution of the first time delay.  This
    delay represents the difference between the time at which we would expect to
    first observe a node (or edge) and the time at which we actually observed it
    in the dataset. This suggests that new nodes and edges continuously appear in
    the media. }
\end{figure*}

Let us denote by $N_t(i)$ the number of co-occurrences of node $i$ with any
other node at time $t$. The attention for neighbours of $i$ follows a similar
pattern as node $i$, more than compared to the overall trend. So, attention for
people that co-occur rises and falls together, hinting at some underlying
commonality. One possibility is that people occur mostly in regard to a specific
issue. The attention of that issue then presumably correlates with the attention
of the people playing a role in that issue, resulting in similar patterns of
attention.

\begin{table*}
  \begin{tabular}{ll|llll}
    & Model & $\alpha - \beta \log |t|$ & $\alpha |t|^{-\beta}$ & $\alpha
    e^{-\beta t}$ & $\alpha e^{-\beta |t|} |t|^{-\gamma}$ \\
    \hline \hline
    \multirow{4}{*}{Growth} & $\alpha$ & $0.011 \pm 6.9 \cdot 10^{-5}$ & $0.031
    \pm 3.2 \cdot 10^{-4}$ & $0.0056 \pm 6.3 \cdot 10^{-5}$ & $0.0233 \pm 1.9 \cdot 10^{-4}$ \\
    & $\beta$ & $0.0013 \pm 9.8 \cdot 10^{-6}$ & $0.49 \pm 2.3 \cdot 10^{-3}$ &
    $0.0019 \pm 3.1 \cdot 10^{-5}$ & $6.6 \cdot 10^{-4} \pm 1.2 \cdot 10^{-5} $ \\
    & $\gamma$ & $-$ & $-$ & $-$ & $0.36 \pm \cdot 2.3 \cdot 10^{-3}$ \\
    & $\Delta \AIC$ & $3\,117$ & $3\,045$ & $4\,883$ & $-$ \\
    \hline
    \multirow{4}{*}{Decay} & $\alpha$ & $0.014 \pm 8.1 \cdot 10^{-5}$ & $0.037
    \pm 3.8 \cdot 10^{-4}$ & $0.0067 \pm 7.1 \cdot 10^{-5}$ & $0.028 \pm 2.3
    \cdot 10^{-4}$ \\
    & $\beta$ & $0.0017 \pm 1.1 \cdot 10^{-5}$ & $0.46 \pm 2.1 \cdot 10^{-3}$ &
    $0.0015 \pm 2.3 \cdot 10^{-5}$ & $5.5 \cdot 10^{-4} \pm 9.8 \cdot 10^{-6}$ \\
    & $\gamma$ & $-$ & $-$ & $-$ & $0.34 \pm 2.2 \cdot 10^{-3}$ \\
    & $\Delta \AIC$ & $2\,559$ & $2\,986$ & $4\,820$ & $-$ \\
  \end{tabular}
  \caption{Growth and decay estimates and model performance. The AIC differences
    are high, and show that the powerlaw with exponential cut-off $\alpha
    e^{-\beta |t|} |t|^{-\gamma}$ is the best model.}
  \label{tab:growth_and_decay}
\end{table*}

Let $t_p(i) = \operatorname{arg\,max}_t N_t(i)$ be the peak of the number of
co-occurrences of node $i$ with any other node (if there are multiple such times
the first is used). We normalise the time series, such that the peak is centred
at $0$ with a value of $1$, and denote the average of these time series by
$\tilde{N}_t$ (see Fig.~\ref{fig:growth_decay_attention}). Hence $\tilde{N}_0 =
1$, and we are interested in how $\tilde{N}_t$ grows for $t < \SI{0}{\day}$ and
decays for $t > \SI{0}{\day}$. \citet{crane_robust_2008} proposed that
$\tilde{N}_t \sim |t|^{-\beta}$, where different exponents $\beta$ would
correspond to different universal classes~\cite{crane_robust_2008}.
However,~\citet{leskovec_meme-tracking_2009} found that $\tilde{N}_t \sim -
\beta \log |t|$, which is unrealistic for large times since $- \log |t| <
\SI{0}{\day}$ for sufficiently large $t$. Alternatively, if the rate of
growth/decay would be constant over time, we would expect to see exponential
growth and decay $\tilde{N}_t \sim e^{-\beta |t|}$. 

However, we find that none of these satisfactorily model the growth and decay of
attention (we use non-linear least squares for fitting). The logarithmic and
exponential functions grow/decay too slowly at a small time (around $10$--$30$
days), while the powerlaw poorly fits the dynamics for larger times. Given that
the powerlaw fits the short time scale relatively well, and the longer time
scales are better fit by an exponential function, a powerlaw with exponential
cut-off $\tilde{N}_t \sim e^{-\beta |t|} |t|^{-\gamma}$ seems a reasonable
alternative.  Indeed, this functional form accurately captures the growth and
decay in our data. For determining which model is a better fit, we use Akaike's
Information Criterion (AIC) values, for which a lower value indicates a better
fit~\cite{burnham_model_2013}. We only give the AIC values relative to the
minimum, which clearly demonstrates that the powerlaw with exponential cut-off
is the best candidate (see Table~\ref{tab:growth_and_decay} for results).
Obviously, the powerlaw diverges for $|t| \to \SI{0}{\day}$ so that it seems as
if the attention is going to diverge, even though we know it is not (because
$\tilde{N}_0 = 1$), and the fitting is done for $|t| > \SI{0}{\day}$.

Although there is a large degree of symmetry---which~\citet{crane_robust_2008}
see as characteristic of endogenous dynamics---we do find different exponents
for the growth and decay (see Table~\ref{tab:growth_and_decay}). In particular,
the decay seems slower than the growth at short time scales, and nodes tend to
occur more frequently after their peak than before their peak on longer time
scales.  This contrasts with the findings
of~\citet{leskovec_meme-tracking_2009}, where the decay was faster than the
growth, and attention was lower after the peak than before the peak.

Let $t_s(i,j)$ be the time of the $s^\text{th}$ co-occurrence between node $i$
and $j$.  The inter-event time can be denoted by $\delta_s(i,j) = t_{s+1}(i,j) -
t_{s}(i,j)$, which can possibly be $0$ if two (or more) co-occurrences happened
at the same day. Similarly for nodes, we denote by $t_s(i)$ the time of the
$s^\text{th}$ co-occurrence of $i$ with another node, and by $\delta_s(i) =
t_{s+1}(i) - t_s(i)$ the inter-event time. If events happen at a constant rate,
we would expect an exponential distribution of inter-event times. If events
follow a bursty patter, we expect to find a powerlaw inter-event time
distribution, often observed in other settings~\cite{malmgren_poissonian_2008}.
We find that although the probability of inter-event times decays quite fast at
a short time scale, the inter-event times for a longer time scale follow an
exponential distribution (Fig.~\ref{fig:iets_distr}). Altogether, a powerlaw
with exponential cutoff $p(x) \sim x^{-\alpha} e^{-\beta x}$ provides a better
fit than a powerlaw or exponential distribution (log-likelihood ratios $6.3
\cdot 10^4$ and $3 \cdot 10^4$ respectively). We employed maximum likelihood
estimation (MLE) for the parameters as recommended
by~\citet{clauset_power-law_2009}. For the edges, we find coefficients $\alpha
\approx 1.003 \pm 2.6 \cdot 10^{-3}$ and $\beta \approx 0.00151 \pm 1.1 \cdot
10^{-5}$, while for the nodes the distribution is slightly less skewed, but
decays slightly faster, with $\alpha = 1.045 \pm 3.7 \cdot 10^{-3} $ and $\beta
= 0.00244 \pm 2.3 \cdot 10^{-5}$, both with a lower cutoff of $x_{\text{min}} =
\SI{10}{\day}$. Indeed, for even shorter time intervals, the decay is very fast,
and using lower $x_{\text{min}}$ gives poorer fits.

One explanation is that if somebody is involved in an issue, his co-occurrence
shows a bursty pattern at this shorter time scale, but that issues appear
at a steady rate over a longer time scale. This could perhaps be
modelled as a cascading Poisson process~\cite{malmgren_poissonian_2008}, where
the rise of an issue triggers the individual events of co-occurrence.

We denote by $\Delta(i,j) = \max t_s(i,j) - \min t_s(i,j)$ the lifetime of an
edge, and similarly by $\Delta(i)$ the lifetime of a node
(Fig.~\ref{fig:lifetime_distr}). Most edges have a very low lifetime of only a
single day, and the probability to have a larger lifetime quickly decreases.
Nonetheless, after an initial rapid decay, the probability decays much slower.
So, besides the more volatile short term links, there are also many long term
stable links. The node lifetimes are rather unusually distributed. Although
there are nodes that have a lifetime that is quite short, nodes tend to have a
longer lifetime, only cutoff by the duration of the dataset. The lifetime of
nodes can thus be long, and can easily run in the decades.

Our data start in about $2004$, but of course some people already appeared in
the media earlier than that. That does not imply we will immediately observe
them the first day (17 January 2004 to be precise), and it may take days, weeks
or perhaps even months before we first observe someone. However, if their
process of occurring in the media is stationary, we can calculate the expected
time until we first observe them, based on the inter-event time distribution:
$\langle t_0(i) \rangle = \frac{\langle \delta_s(i)^2 \rangle}{2 \langle
\delta_s(i) \rangle}$ (and similarly so for
edges)~\cite{lawler_introduction_1995}. We call the difference $\Delta t_0(i) =
t_0(i) - \langle t_0(i) \rangle$ the ``first time delay'', and if people already
appeared in the news before $2004$, we should find that $\Delta t_0(i) \approx
\SI{0}{\day}$ for those people. If $\Delta t_0(i)$ is much larger for somebody,
it becomes unlikely that this person appeared in the news prior to $2004$ (still
assuming the process is stationary). The distribution of the first time delay is
displayed in Fig.~\ref{fig:first_time_delay_distr}.  First, there is a clear
peak around $\Delta t_0(i) = 0$, suggesting that these nodes indeed already
appeared in the news prior to $2004$. The fast decay that is visible on the
negative part, does not show at the positive part. This implies that many nodes
and edges are first observed much later than expected.  Hence, new nodes and
edges continuously appear in the news. This is especially prominent for the
edges, which show an almost uniform distribution between $250$--$2\,000$ days of
difference, whereas the distribution for the nodes decays more gradually.

\section{Conclusion}

Online social media show signs of exogenous shocks, bursty dynamics and
exponential lifetime distributions. We have shown here that traditional media
work differently, with different regimes operating at two different time
scales. There is a short time scale at which links have short lifetime,
attention decays quickly and there are indications of burstiness. However, at a
longer time scale nodes and links have a relatively long lifetime, attention
decays slower, and inter-event times decay exponentially. We believe that this
might be modelled as a cascading Poisson
process~\cite{malmgren_poissonian_2008}. Issues would arise at a steady rate,
triggering individual events of co-occurrence. The longer time scale would be
related to the slower dynamics of how issues appear and disappear, while the
faster dynamics at a shorter time scale would be due to the individual events
within an issue. We aim to further pursue this idea in future research.

\bibliographystyle{iccsa}
\bibliography{media_dynamics}

\end{document}